\documentclass[a4paper,12pt]{article}
\pdfoutput=1 

\usepackage{jcappub} 
\usepackage{subfigure}
\usepackage[T1]{fontenc} 
\usepackage{lipsum}
\usepackage{amsmath, multirow, amssymb, wasysym, gensymb}
\usepackage{cleveref}

\usepackage{graphicx}
\usepackage{bm}
\usepackage{times}
\usepackage{slashed}
\usepackage{color}
\usepackage{epsfig}\usepackage{dcolumn}
\usepackage{color}
\def\br{\begin{eqnarray}}
\def\er{\end{eqnarray}}
\def\be{\begin{equation}}
\def\ee{\end{equation}}

\usepackage{indentfirst}

\title{Lower Mass Bound on the $W^\prime$ mass via Neutrinoless Double Beta Decay in a 3-3-1 Model}

\author[a,b]{A. C. O. Santos,}
\author[a]{P. Vasconcelos}

\affiliation[a]{Departamento de F\'isica, Universidade Federal da Para\'iba,
Caixa Postal 5008, 58051-970, Jo\~ao Pessoa, PB, Brazil} 
\affiliation[b]{Centre for Cosmology, Particle Physics and Phenomenology (CP3), Universit\'e catholique de Louvain, B-1348, Louvain-la-Neuve, Belgium
}

\emailAdd{antonio\_santos@fisica.ufpb.br}
\emailAdd{pablosje@gmail.com}

\abstract{
The discovery of neutrino masses has raised the importance of studies in the context of neutrinoless double beta decay ($0\nu\beta\beta $), which constitutes a landmark for lepton number violation (LNV). The standard interpretation is that the light massive neutrinos, that we observed oscillating in terrestrial experiments, mediate double beta decay. In the minimal 3-3-1 model (3-3-1M), object of our study, there is an additional contribution that stems from the mixing between a new charged vector boson, $W^{\prime}$, and the Standard Model W boson. Even after setting this mixing to be very small, we show that tight constraints arise from the non-observation of ($0\nu\beta\beta $). Indeed, we derive bounds on the mass of the $W^{\prime}$ gauge boson that might exceed those from collider probes, and most importantly push the scale of symmetry breaking beyond its validity, leading to an exclusion bound for the minimal 3-3-1 model. 
}

\begin{document}
\maketitle
\flushbottom

\section{Introduction}
\label{sec:intro}

\begin{figure}[ht]
\centering
\includegraphics[scale=0.7]{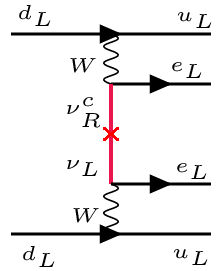}
\caption{Canonical $0\nu\beta\beta $ due to light massive neutrinos.}
\label{fig1}
\end{figure}

The Standard Model (SM) has thrived after a multitude of precision tests at low and high energy scales \cite{Olive:2016xmw}. However, the existence of neutrino masses is one of the main motivations for physics beyond the SM \cite{Rodejohann:2011mu}. If neutrinos are majorana particles, neutrinoless double beta decay ($0\nu\beta\beta $) should occur. Double beta decay is the transition of a nucleus with mass and atomic number $A$ and $Z$ to a nucleus with $A$ and $Z+2$, accompanied by the emission of two electrons only. Its possible discovery will represent an irrefutable proof of Lepton Number Violation $LNV$. The standard diagram that leads to such lepton number violation process is exhibited in Fig.\ref{fig1}.

At present, a lot of experiments using different isotopes and techniques are operating or under development in the search for neutrinoless double beta decay (see \cite{Pas:2015eia}). In this work we investigate the implications of the non-observation of neutrinoless double beta decay in the context of 3-3-1 models \cite{Pisano:1991ee,Foot:1992rh,Frampton:1992wt,Foot:1994ym,Hoang:1995vq}. 3-3-1 models are plausible extensions of the SM where fermions are placed in the fundamental or adjoint representation of $SU(3)$. Generally these models while being consistent with colliders data \cite{Alves:2011kc,Alves:2012yp,Caetano:2013nya}, nicely explain why we have three fermion generations \cite{Pisano:1991ee,Dias:2004dc,Dias:2004wk}, accommodate neutrino masses \cite{Queiroz:2010rj,Caetano:2012qc,Cogollo:2009yi}, and may feature several dark matter candidates \cite{Mizukoshi:2010ky,Alvares:2012qv,Queiroz:2013lca,Profumo:2013sca,Kelso:2013nwa,Queiroz:2013zva,Dong:2014wsa,Kelso:2014qka,Cogollo:2014jia,Alves:2016fqe,deS.Pires:2010fu} to the dark matter problem \cite{Queiroz:2016sxf,Kavanagh:2017hcl}. Moreover, they could also address low energy anomalies \cite{Cogollo:2012ek,Kelso:2013zfa,Cogollo:2013mga,Queiroz:2014zfa,Allanach:2015gkd,Queiroz:2016gif,Altmannshofer:2016jzy,Lindner:2016bgg} and might feature interesting astrophysical \cite{Hooper:2011aj,Hooper:2012sr,Kelso:2013paa,Queiroz:2014yna,Alves:2014yha,Baring:2015sza,Mambrini:2015sia,Profumo:2016idl,Queiroz:2016awc,Campos:2017odj,Arcadi:2017vis} and collider phenomenology via the presence of exotic gauge bosons which are popularly used in many different contexts \cite{Alves:2013tqa,Alves:2015mua,Alves:2015pea,Alves:2016bib,Alves:2016cqf,Gonzalez-Morales:2014eaa,Patra:2015bga,Lindner:2016lxq,Lindner:2016lpp,Altmannshofer:2016jzy,Queiroz:2016qmc,Klasen:2016qux,Arcadi:2017jqd,Arcadi:2017vis,Arcadi:2017atc,Arcadi:2017kky,Campos:2017dgc,Arcadi:2017hfi}. See \cite{Doff:2015nru,Ferreira:2016uao,Rodriguez:2016cgr,Pires:2016vek,Borges:2016nne,Dong:2017ayu,Ferreira:2017mvp,Cogollo:2017foz} for many other exciting phenomenological studies.\\

In this work, we focus our attention on the minimal 3-3-1 model (3-3-1M), one of the possible versions of 3-3-1 models, in which the electric operator takes the form $ Q=T_{3} + \beta T_{8} + X $ considering $T_{i}$ as $\frac{\lambda^{a}}{2}$ (Gell-Mann matrices with $a=1,...,8$), $\beta=-\sqrt{3}$ for $SU(3)_{L}$, see explanation below, and $X$ as $U(1)_{X}$ charge, where no new leptons are evoked (right-handed neutrinos as well), thus featuring a minimal fermion content. Since we are dealing with an extended gauge group, there will be additional gauge bosons, one of them being singly charged, a $W^\prime$, due to the chosen charge operator parameter. This new gauge boson might induce the neutrinoless double beta decay if it mixes with the W boson. In order to successfully explain neutrino masses in this model without the use of non-renormalizable operators \cite{Queiroz:2010rj}, a scalar sextet should be introduced \cite{Montero:2001ts,Montero:2001tq}. This scalar sextet embeds the scalar triplet model, often used in the context of neutrino masses \cite{Mohapatra:1999zr,Gu:2006wj,Akhmedov:2006de,Reig:2016ewy}. The first component of the scalar sextet is a neutral field which is often assumed to yield a null vacuum expectation value ($vev$). In this work we assume otherwise. If such scalar acquires a non-vanishing   $vev$ then the $W^\prime$ mixes with the SM W boson, giving rise to new contributions to the neutrinoless double beta decay. This scenario has been investigated before in the context of majoron in \cite{Montero:1999mc,Montero:1999su,Montero:2000ar,Montero:2001ji}. Here we revisit the implications in a more general setting in perspective with other existing colliders constraints. We will ignore contributions arising from the scalar fields because they yield less restrictive bounds \cite{Montero:1999su} and because a bound on the $W^\prime$ mass already implied into a lower mass bound on the entire particle spectrum of model, since the $W^\prime$ mass is directly connected to the scale of symmetry breaking of the model. It is worth nothing to mention that in this work our space parameter provides a very small $Z-Z^{\prime}$ mixing and does not bring relevant constraints coming from $STU$ parameters \citep{Montero:2001ji,Dong:2014bha}.  

We will show that if the $W^\prime$ does feature a mixing with the SM $W$ boson then limits stemming from neutrinoless double beta decay supersedes LHC probes, highlighting the importance of multiple new physics searches probes.

The paper is organized as follows: In section II we describe the model and explain how the $W^\prime - W$ mass mixing can be generated. In section III we derive the contributions to neutrinoless double beta decay. In section IV we draw our conclusions.

\section{The model}
\label{sec:model}

In the minimal 3-3-1 model leptons are arranged as,

\begin{equation}
L^{\ell}  = \left( 
\begin{array}{c}
\nu_{\ell}\\
\ell\\
\ell^c\\
\end{array}\right)_{L} \sim (3,0),
\end{equation}
under $SU(3)_L \otimes U(1)_N$, with $\ell=e,\mu,\tau$, representing the three known generations. Notice that the third component of the lepton triplet is the right-handed lepton. The quarks are placed as follows, 

\begin{equation}
Q_{1L}= \left( 
\begin{array}{c}
u_1 \\
d_1 \\
J_1 \\
\end{array}\right)_{L}\sim (3,2/3),\quad  
Q_{iL}=\left( 
\begin{array}{c}
d_i \\
-u_i\\
J_i \\
\end{array}\right)_L\sim (\bar{3},-1/3),\quad i=2,3.
\end{equation}
Here $u_{aL}$ and $d_{aL}$ ($a=1,2,3$) correspond respectively to flavors $u$, $c$, $t$, and $d$, $s$, $b$ of the Standard Model (SM) quarks. The first generation transform as a triplet, while the second and third generations transform as anti-triplets under $SU(3)_L$. In addition to the left-handed field we have right-handed quarks as singlets under $SU(3)_L$: $u_{aR}\sim (1,2/3)$, $d_{aR}\sim (1,-1/3)$, $J_{1R}\sim (1,5/3)$, $J_{iR}\sim (1,-4/3)$, with $a=1,2,3$ and $i=2,3$. Being $J_1$ and $J_i$ exotic quarks predicted by the model with electric charges of $5/3$ and $-4/3$, respectively. These exotic quarks are also known as leptoquarks in the literature \cite{Queiroz:2014pra,Allanach:2015ria}.
 
The mass generation mechanism rely firstly on the presence of three scalar triplets, namely

\begin{equation}
\eta=\left( 
\begin{array}{c}
\eta^0\\
\eta_1^-\\
\eta_2^+\\
\end{array}\right),\rho=\left( 
\begin{array}{c}
\rho^+\\
\rho^0\\
\rho^{++}\\
\end{array}\right),\chi=\left( 
\begin{array}{c}
\chi^-\\
\chi^{--}\\
\chi^0\\
\end{array}\right),
\end{equation}
with them transforming as $(3,0)$, $(3,1)$ and $(3,-1)$ under $SU(3)_L \otimes U(1)_N$, respectively.
These scalar triplets couple with the fermions fields through the following Yukawa lagrangian
\begin{eqnarray}
-{\cal L}_Y &=&  \frac{1}{2}\sum_{\ell\ell^{\prime}}G^{\ell\ell^{\prime}} \epsilon^{lmn}\left[ \overline{(L^{\ell}_{l})^{c}}L^{\ell^{\prime}}_{m}\eta_{n} \right] \nonumber  \\
&+& G^{u}_{1a}\bar{Q}_{1L} \eta u_{aR} + G^{d}_{1a}\bar{Q}_{1L}\rho d_{aR} + G^{J1}_{11}\bar{Q}_{1L}\chi J_{1R}          \nonumber  \\
&+& G^{s}_{2a}\bar{Q}_{2L}\eta^{*} d_{aR} + G^{c}_{2a}\bar{Q}_{2L}\rho^{*} u_{aR}+ G^{J2}_{2i}\bar{Q}_{2L}\chi^{*} J_{iR} \nonumber  \\
&+& G^{b}_{3a}\bar{Q}_{3L}\eta^{*} d_{aR} + G^{t}_{3a}\bar{Q}_{3L}\rho^{*} u_{aR}+ G^{J3}_{3i}\bar{Q}_{3L}\chi^{*} J_{iR} + h.c.,
\label{e10}
\end{eqnarray} 
with $\ell,\ell^{\prime}=e,\mu,\tau$. Eq.\ref{e10} is sufficient to generate masses to charged leptons and quarks but neutrinos remain massless. In order to obtain non-zero neutrino masses, a scalar sextet should be introduced,

\begin{equation}
S=\left( 
\begin{array}{ccc}
\sigma _1^0 & \frac{h_2^{-}}{\sqrt2} & \frac{h_1^{+}}{\sqrt2} \\ 
\frac{h_2^{-}}{\sqrt2} & H_1^{--} & \frac{\sigma _2^0}{\sqrt2} \\ 
\frac{h_1^{+}}{\sqrt2} & \frac{\sigma _2^0}{\sqrt2} & H_2^{++}
\end{array}
\right) \sim \left( {\bf 6},{\bf 0}\right).
\label{s}
\end{equation}

The presence of such scalar sextet gives rise to the yukawa term,
\begin{equation}
{\cal L}^{S}_{Y}=-\frac{1}{2}\sum_{\ell\ell^{\prime}} G \overline{(L^{\ell})^c}S^{*}L^{\ell^{\prime}},
\label{e4}
\end{equation}
with $L^c = C\bar{L}^T$, being $C$ the charge conjugate matrix.

Expanding Eq.\ref{e4} we have explicitly
\begin{eqnarray}
{\cal L}^{S}_{Y}& \supset & -\frac{1}{2}\sum_{\ell\ell^{\prime}}G_{\ell\ell^{\prime}} [ \overline{(\nu_{\ell L})^c} \nu_{\ell^{\prime} L} \sigma_1^0 + \overline{ (\ell_{L})^c}\ell^{\prime}_{L} H_1^{++} + \overline{ \ell_{R}}(\ell^{\prime c})_{L}H_2^{--}\nonumber \\
&& + \left( \overline{(\ell_{L})^c}\nu_{\ell^{\prime} L} +\overline{(\nu_{\ell L})^c}\ell^{\prime}_{L} \right) \frac{h_2^{+}}{\sqrt{2}}
+ \left(\overline{\ell_{R}} \nu_{\ell^{\prime} L} + \overline{(\nu_{\ell L})^{c}}(\ell^{\prime c})_{L} \right) \frac{h_1^{-}}{\sqrt{2}}  \nonumber \\
&& + \left(\overline{\ell_{R}}\ell^{\prime}_{L} + \overline{(\ell_{L})^{c}}(\ell^{\prime}_{R})^{c} \right) \frac{\sigma_{2}^{0}}{\sqrt{{2}}} ]+h.c.
\label{yul}
\end{eqnarray}

Notice that a new contribution to the charged lepton masses arise if $v_{\sigma_2} \neq 0$. Moreover, neutrinos remain massless unless $\langle \sigma^0_1\rangle \neq 0$. Typically, this scalar sextet is absent in 3-3-1 studies, and consequently neutrino masses are not addressed. In this work, we discuss a more general setting, where the scalar sextet is present and 
$v_{\sigma_{1,2}} \neq 0$. In this setup, the spontaneous symmetry breaking occurs as follows: firstly, $\chi^0$ develops a $vev$ ($v_{\chi}$), breaking $SU(3)_L\otimes U(1)_N$ into $SU(2)\otimes U(1)_Y$. Later $\rho^0$ and $\eta^0$ acquire a non zero $vev$ with $v_{\rho}\sim v_{\eta}$, breaking $SU(2)\otimes U(1)_Y$ into $U(1)_{QED}$, i.e. the electromagnetism gauge group. 

This symmetry breaking pattern generates mass to all SM fermions and gauge bosons. We highlight that due to the enlarged gauge group, 3-3-1 models feature five new bosons. In the minimal 3-3-1 model they are identified as $W^{\prime \pm}$, $U^{\pm \pm}$, $Z^\prime$. An important result of the spontaneous symmetry breaking is the rising of a mass mixing between the W and $W^\prime$ bosons that yields the mass mixing matrix, 

\begin{equation}
\left(\begin{array}{cc}
 W^+_\mu & W^{\prime +}_\mu
\end{array}
\right)
\left( 
\begin{array}{cc}
M^2_W & M^2_{WW^{\prime}} \\
\newline \\
M^2_{WW^{\prime}} & M^2_{W^\prime}
\end{array}
\right)
\left(
\begin{array}{c}
W^{\mu -}\\ 
\newline \\
W^{\prime \mu -}
\end{array}
\right),
\label{mm}
\end{equation} where

\begin{equation}
M^2_{WW^{\prime}}=\frac{g^2}{2}(2v_{\sigma_1}v_{\sigma_2}),
\end{equation} 
$g=e/s_W$, and

\begin{eqnarray} 
M^2_W & =& \frac{g^2}{2}\left( v^2_\eta + v^2_\rho+ v^2_{\sigma_2}+v_{\sigma_{1}}^2\right),\nonumber\\
\newline \\
M^2_{W^\prime} & =& \frac{g^2}{2}\left(v^2_\eta+v^2_\chi+v^2_{\sigma_2}+v^2_{\sigma_1}\right). \nonumber
\label{Wmass}
\end{eqnarray}

If we take $v_{\sigma_{1,2}}=0$, then $M^2_{WW^{\prime}}=0$, and consequently $v_\eta^2+v_\rho^2 = v^2$, where $v=175\,$GeV. In this way, the W mass is correctly obtained. After diagonalization we find two mass eigenstates $W_1^+$ and $W_2^+$ with,

\begin{equation}
M_{W_{1,2}}=\frac{1}{2}\left[(M^2_W+M^2_{W^\prime}) 
\pm\left((M^2_W-M^2_{W^\prime})^2+4 M^4_{WW^{\prime}} \right)^{1/2}\right],
\label{massnow}
\end{equation}which are related to the mass eigenvectors via,

\begin{equation}
\left(
\begin{array}{c}
W^+ \\ W^{\prime +}
\end{array}
\right)=\left(
\begin{array}{cc}
c_\theta & -s_\theta\\
s_\theta & c_\theta
\end{array}
\right)
\left(
\begin{array}{c}
W^+_{1} \\ W^+_{2}
\end{array}
\right),
\label{me}
\end{equation}
with 
\begin{equation}
\tan2\theta=\frac{-2M^2_{WW^{\prime}}}{(M^2_W-M^2_{W^\prime})}.
\end{equation}

Considering the limit $M^2_{WW^{\prime}} \rightarrow 0$, we get $M_{W_1}=M_W$ and $M_{W_2}=M_{W^\prime}$. The charged currents associated to these gauge bosons are,

\begin{eqnarray}
\mathcal{L}_{cc} &\supset & \frac{g}{\sqrt{2}} \left[ \bar{u}_L \gamma^\mu V_{CKM} d_L - \bar{\nu}_L \gamma^\mu (U_\nu V_{\ell}) l_L \right] W^+_\mu \nonumber\\
 &+& \frac{g}{\sqrt{2}} \left[\bar{l^c}_L \gamma^\mu (V_l\, U_\nu^\dagger) \nu_{L} \right] W^{\prime +}_{\mu} + h.c.,
\end{eqnarray}
where $V_l$ and $U_\nu$ are the mixing matrices for the charged leptons and neutrinos, with $\nu_{L}=(\nu_{eL},\nu_{\mu L},\nu_{\tau L})$ and $l_{L}=(e_{L},\mu_{L},\tau_{L})$ . There is also a term involving the doubly charged gauge boson, but it is not relevant for our reasoning \cite{Montero:2000ar}. 

We have gathered all ingredients to now discuss the neutrinoless double beta decay in the minimal 3-3-1 model.

\section{The neutrinoless double beta decay}
\label{sec:bb0n}

Neutrinoless double beta decay is a landmark process in particle physics. It is defined as the transition of a nucleus into another nucleus with an atomic number larger by two units, and the emission of two electrons only,

\begin{equation}
(A,Z) \rightarrow (A,Z+2) +2 e^-.
\end{equation}

Since there are no leptons in the initial state, but two in the final state, the observation of neutrinoless double beta decay constitute an irrefutable proof that lepton number is violated by nature. In the past decades, there has been a substantial improvement on the bound over the half-life of the neutrinoless double beta decay \cite{DellOro:2016tmg}. These bound can be later translated into upper limits on the effective majorana mass defined as,

\begin{equation}
m_{\beta \beta} = \left| \sum_{i=e,\mu,\tau}  U_{ei} ^{2} m_{\nu_i}\right| .
\end{equation}

\begin{figure}
\centering
\includegraphics[scale=0.6]{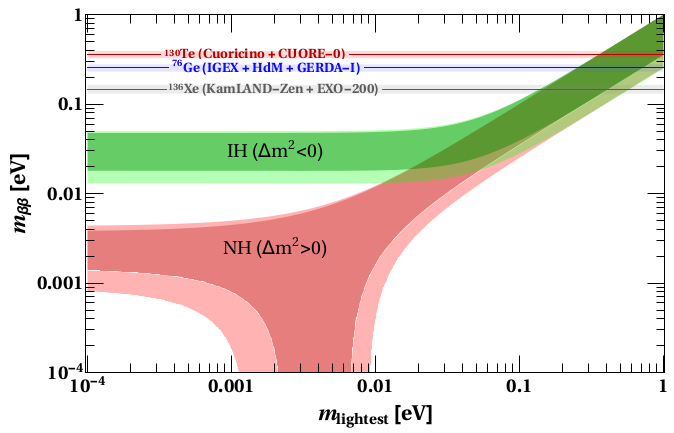}
\caption{Updated prediction on $m_{\beta\beta}$ from neutrino oscillations as a function of the lightest neutrino mass according to the current neutrino data. In the figure, both normal and inverted hierarchy schemes for neutrino masses are displayed. Figure taken from \cite{DellOro:2016tmg}. In the figure is also visible the current limit from KamLAND-Zen.}
\label{fig2new}
\end{figure}

The effective majorana mass term grows inversely with the half-life. A stronger lower bound on the half-life of the neutrinoless double beta decay implies a stronger bound on the effective majorana mass term (see Fig.\ref{fig2new}). The strongest bound today comes from KamLAND-Zen \cite{Asakura:2014lma}, $t_{1/2} > 2.6 \times 10^{25}$yrs, implying that,

\begin{equation}
m_{\beta \beta}^{current} < 0.15 {\rm eV}.
\end{equation}

In the foreseeable future, CUORE is expected to achieve $^{130}Te > 9.5 \times 10^{25}$yrs  \cite{DellOro:2016tmg} which translates into,
\begin{equation}
m_{\beta \beta}^{CUORE} < 0.073 {\rm eV}.
\end{equation}

A projected limit further into the future with nEXO of $t_{1/2} > \times 10^{27}$~yr for $^{136}Xe$ is expected, which then would yield,

\begin{equation}
m_{\beta \beta}^{nEXO} < 0.01 {\rm eV}.
\end{equation}

Now from  the current and projected experimental sensitivity we will approach the theoretical aspects of this observable. That said, the amplitude in Fig.\ref{fig1} is proportional to,

\begin{equation}
A_1\propto\frac{g^4   m_{\beta\beta}}{M^4_W\langle p^2\rangle}\,
c^4_\theta
=\frac{32G^2_F m_{\beta\beta} }{\langle p^2\rangle}\,c^4_\theta,
\label{a1}
\end{equation}
where $\langle p^2\rangle$ is the average 
of the four-momentum transfer squared, which is approximately
$(100\,\mbox{MeV})^2$.

\begin{figure}[ht]
\centering
\includegraphics[scale=0.8]{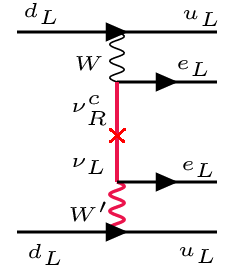}
\caption{Novel 331 contribution for double beta decay due to light massive neutrinos.}
\label{fig2}
\end{figure}

Moreover, one should observe that the amplitude for the diagram in Fig.\ref{fig2} is proportional to,

\begin{equation}
A_2\propto 
32G^2_F\,\left(\frac{M_W}{M_{W^{\prime}}} \right)^2
\frac{c^3_\theta s_\theta}{\sqrt{\langle p^2\rangle}}.
\label{a2}
\end{equation}

Having in mind that $\theta$ should be small in order not to alter the SM W properties, then $A_2/A_1 < 1$. This results into,

\begin{equation}
\frac{A_2}{A_1}=\left( \frac{M_W}{M_{W^\prime}}\right)^2
\frac{\sqrt{\langle p^2\rangle}}{m_{\beta\beta}}\tan\theta < 1.
\label{a2a1}
\end{equation}

Therefore we get the lower mass bound, 

\begin{equation}
M_{W^\prime} > 2.540\, {\rm TeV}\left(\frac{0.1\, {\rm eV}}{m_{\beta\beta}}\right)^{1/2} \times \left(\frac{\tan\theta}{10^{-6}}\right)^{1/2}.
\label{c1}
\end{equation}

\begin{figure}[ht]
\centering
\includegraphics[scale=0.7]{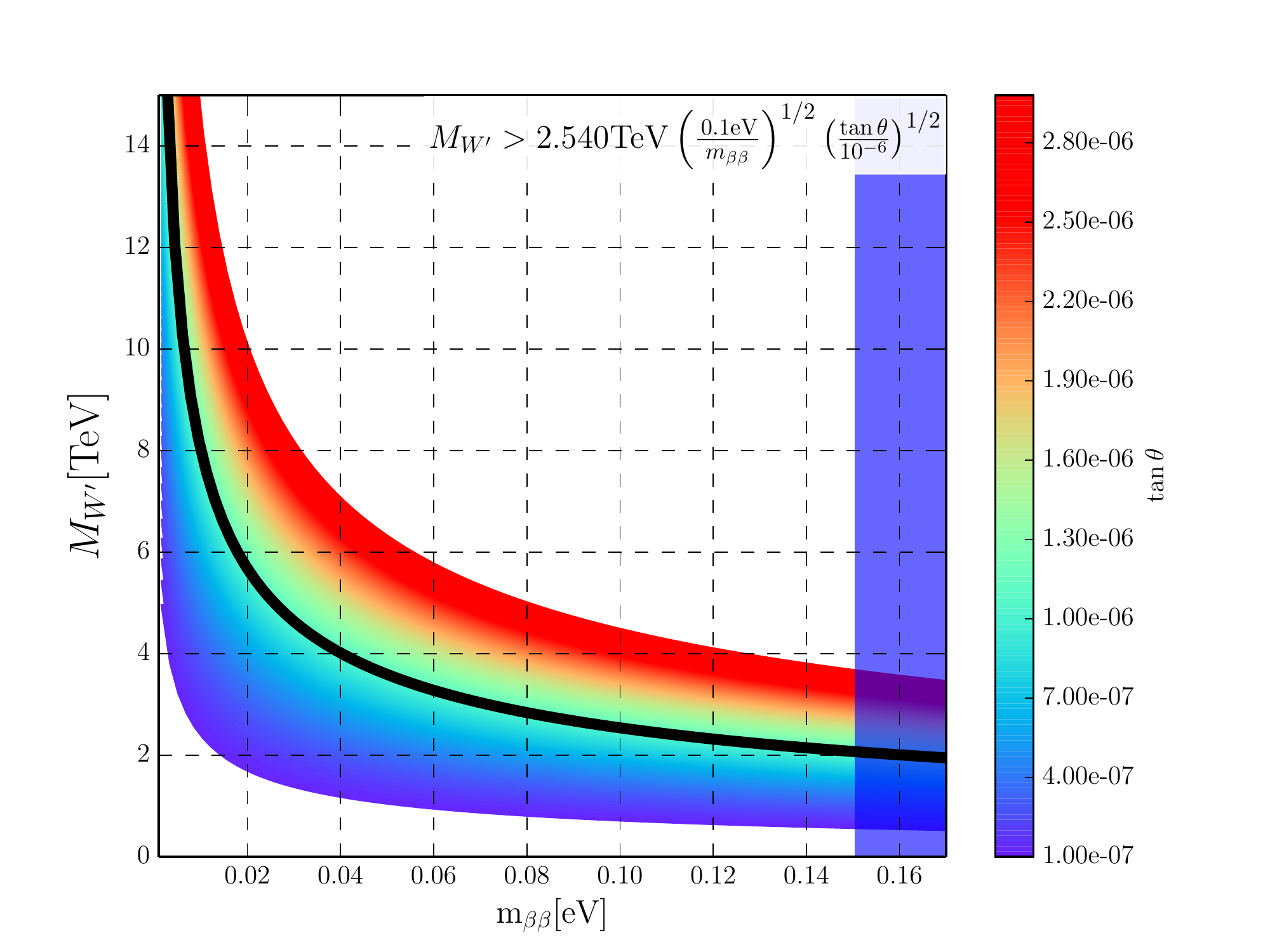}
\caption{Temperature plot of the lower mass bound on the $W^\prime$ mass as function of $m_{\beta\beta}$ and $\tan \theta$. The black curve is drawn simply to guide the eye and check that indeed for $\tan\theta \sim 10^{-6}$ and $m_{\beta\beta}\sim 0.1$eV, we find the lower mass bound $M_{W^\prime} \gtrsim 2.5$TeV. }
\label{fig3}
\end{figure}

In Fig.\ref{fig3} we exhibited this lower mass bound on the $W^\prime$ mass. It is quite visible that even for small $\tan\theta$ the bound on the $W^\prime$ is rather strong. 

However, \ref{c1} is not very useful to us because the $W^\prime$ mass also appears in $\tan\theta$. Therefore, these quantities are not independent they, are strongly correlated. Having in mind that we are working in the regimen where $\theta \ll 1$, then $\tan\theta \sim \theta$ and we can solve for $M_{W^\prime}$ to find,

\begin{equation}
M_{W^\prime} > 50.4\, {\rm TeV}\left(\frac{0.1\, {\rm eV}}{m_{\beta\beta}}\right)^{1/4}  (M^2_{WW^{\prime}})^{1/4}.
\label{lowerbound}
\end{equation}

This equation can give rise to more robust bounds on the $W^\prime$ mass since the parameters that go into it are the majorana effective mass and $M^2_{WW^{\prime}}$ term, that depends on the $vevs$ of the scalar fields $\sigma_1$ and $\sigma_2$. Since the Standard Model $W$ mass does not comes mostly from the $vevs$ of these scalar field as long as they are much smaller than $v_{SM}$, then our choices for $v_{\sigma_1}$ and $v_{\sigma_2}$ are in principle completely arbitrary.

\begin{figure}[ht]
\centering
\includegraphics[scale=0.9]{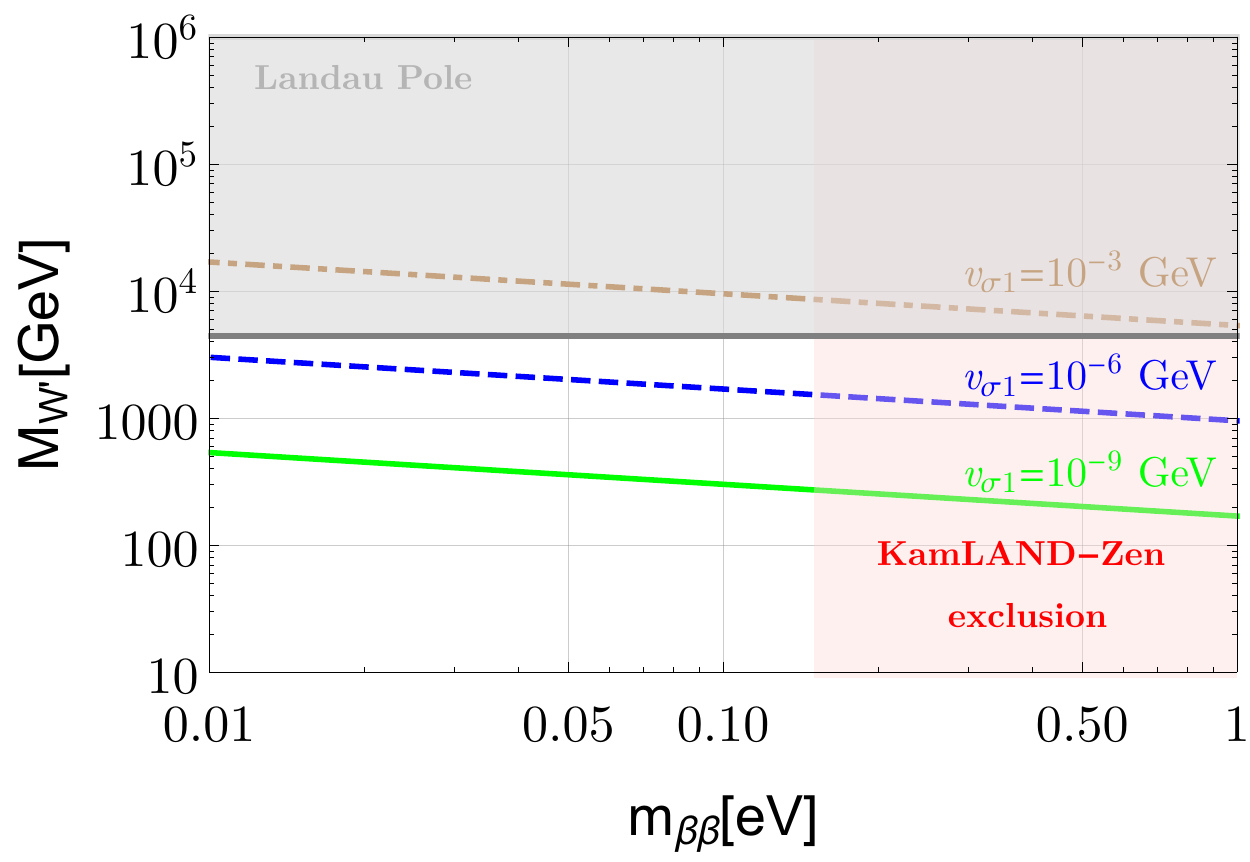}
\caption{Lower mass bound on the $W^\prime$ mass from the non-observation of the neutrinoless double beta decay with different choices for the $vev$ of scalar field $\sigma_1$ which controls the parameter $M^2_{WW^{\prime}}$ in Eq.\ref{lowerbound}. The gray solid area denotes Landau pole upper limit \cite{Dias:2004dc,Martínez2007} and in light red the KamLAND-Zen exclusion \cite{Asakura:2014lma}.}
\label{fig4}
\end{figure}

To have a clear vision of what Eq.\ref{lowerbound} represents, we show the lower mass bounds on the $W^\prime$ mass in Fig.\ref{fig4}. There we select $v_{\sigma2}=3\,$GeV, larger values are ruled out by the $\rho$-parameter \cite{Sharma:1999ct,Montero:1999su}, and vary $v_{\sigma1}$. Notice that, depending on the value adopted for $v_{\sigma1}$, neutrinoless double beta decay might yield very strong constraints on the $W^\prime$ mass. In particular for $m_{\beta\beta} \sim 0.1\,$eV we get a lower mass bound on the $W^\prime$ mass that varies from $ 300\,$GeV to about $ 9.5$\,TeV. The latter is achieved if $v_{\sigma1}=10^{-3}\,$GeV. 

One shortcoming of our lower mass bound on the $W^\prime$ mass is its dependence on the choice for $vevs$. Anyways, these constitute an independent bound on the $W^\prime$ mass which can be much stronger than the limit on the $W^\prime$ arising from colliders searches. Notice that our $W^\prime$ does not couple only to ordinary quarks, therefore it cannot be produced resonantly at the LHC via s-channel processes. Nevertheless, one can use LHC limits on the $Z^\prime$ mass to convert into a lower mass limit on the $W^\prime$ mass, since both masses are dictated by the same quantity, $v_{\chi}$, the scale of symmetry breaking of the 3-3-1 symmetry. Indeed one can find that $M_{W^\prime} \sim 0.32 v_{\chi}$ and $M_{Z^\prime} \sim 0.4 v_{\chi}$. Limits from dilepton searches at the LHC at 14 TeV with $\sim 23 fb^{-1}$ exclude $M_{Z^\prime}  < 4$\,TeV (see Fig.9 of \cite{Coutinho:2013lta}). A projection with $100fb^{-1}$ of data, extrapolating the luminosity effect, one would exclude $M_{Z^\prime}<4.7$\,TeV. These two limits translate into $M_{W^\prime}> 3.2$\,TeV and $M_{W^\prime}> 3.7$\,TeV, respectively. 

Therefore, we can conclude the neutrinoless double beta decay offers a complementary probe to collider physics and depending on choices for $v_{\sigma1}$, it can offer the most restrictive bound on this gauge boson mass. We emphasize that the mixing angle $\theta$ is quite small since it is dictated by $M^2_{WW^{\prime}}/M_{W^\prime}^2$, rendering our conclusions robust.

\section{Validity of the Minimal 3-3-1  Model}

 It has been shown that the 3-3-1M is valid up to energies of $5$~TeV or so due the presence of a Landau pole \cite{Dias:2004dc,Martínez2007}. The collider bounds on the $Z^\prime$ and $W^\prime$ mentioned above already pose a strong tension on the model since these mass limits translate into $v_{\chi} \gtrsim 7$\,TeV and $v_{\chi} \gtrsim 8.2$\,TeV. With the addition of new exotic fermions that can contribute to the renormalization group equations, this tension can be alleviated . As can be seen in Fig.\ref{fig4}, for $m_{\beta\beta} \sim 0.1$\,eV and $v_{\sigma 1}\sim 10^{-6} $\,GeV, the lower mass bound on the $W^\prime$ lies around $1.8$\,TeV, which implies that $v_{\chi} > 3.6$\,TeV. 

This would constitute a strong but valid claim, under the assumptions made. Neutrinoless double beta decay under certain assumptions already excludes the minimal 3-3-1 model, imposing that the scale of symmetry breaking of the model should lie beyond its validity \footnote{for previous bounds on this model see \cite{Diaz:2003dk,Diaz:2004fs,GonzalezSprinberg:2005zd,Ochoa:2005ch,CarcamoHernandez:2005ka,Dias:2005jm,Cabarcas:2008ys,Martinez:2009ik,Cabarcas:2009vb}}. 

These findings are valid for $m_{\beta\beta} \sim 0.1$eV, but if we consider the nEXO sensitivity which is expected to reach $m_{\beta\beta} \sim 0.01$eV, the impact that our study brings to the minimal 3-3-1 model is even more profound. Even for $v_{\sigma 1} \sim 10^{-6}$\,GeV, we would already impose $v_{\chi}\geq 6.5$\,TeV.

In summary, neutrinoless double beta decay offers an orthogonal probe to the minimal 3-3-1 model. If the $vevs$ of the fields in the scalar sextet are sufficiently small neutrinoless double beta decay does not favor the original version of the minimal 3-3-1 model.

\section{Conclusions}

We discussed neutrinoless double beta decay in the context of the minimal 3-3-1 model. It features a minimal fermion content, arguably rendering it more predictive in comparison to other models based on this 3-3-1 gauge group. In the past decades we have observed a significant improvement on the bound of the neutrinoless double beta decay. We exploit this fact to obtain a lower mass bound on the $W^\prime$ boson that arises due to the enlarged gauge group. Our limits are based mostly on the charged current and the $vev$ of a scalar field that generates neutrino masses and induces the $W-W^\prime$ mixing. This mixing leads to a sizable contribution to neutrinoless double beta decay. 

We obtained a fully analytic expression that represents a lower mass bound on the $W^\prime$ mass. Depending on the $vev$ adopted for this scalar field ($\sigma_1$), neutrinoless double beta decay might offer the most restrictive limit on the $W^\prime$ surpassing those of collider probes. Moreover, since the $W^\prime$ mass is directly connected to the scale of symmetry breaking of the model, a lower mass bound on the gauge boson represents also a lower bound on the scale of symmetry breaking. In particular, if $v_{\sigma 1}>0.1$\,GeV, current limits on the half-life of the neutrinoless double beta decay strongly disfavored out the minimal 3-3-1 model.

\acknowledgments

This work was partly supported by the Conselho
Nacional de Pesquisa e Desenvolvimento Cient\'ifico- CNPq, and also by Coordena\c c\~ao de Aperfei\c coamento de Pessoal de N\'ivel Superior (Capes-PDSE-88881.135139/2016-01). 

\bibliographystyle{JHEPfixed}
\bibliography{darkmatter}

\end{document}